\documentclass[aps,prl,preprint,tightenlines,superscriptaddress,showpacs,byrevtex]{revtex4}
\usepackage{graphicx} 
\usepackage{dcolumn}  
\usepackage{graphics} 
\usepackage{epsfig} 
\usepackage{amsmath} 
\usepackage{amssymb} 
\usepackage{textcomp} 

\graphicspath{{ps}}

\begin{document}

\preprint{\vbox{  
                  \hbox{ } 
                  \hbox{ }
                  \hbox{Belle-CONF-1402}
                  \hbox{For the Tau-2014 conference}
}}

\vspace*{15mm}
\title{ \quad\\[0.5cm] Study of Michel parameters in leptonic $\tau$ decays at Belle}


\noaffiliation
\affiliation{University of the Basque Country UPV/EHU, 48080 Bilbao}
\affiliation{Beihang University, Beijing 100191}
\affiliation{University of Bonn, 53115 Bonn}
\affiliation{Budker Institute of Nuclear Physics SB RAS and Novosibirsk State University, Novosibirsk 630090}
\affiliation{Faculty of Mathematics and Physics, Charles University, 121 16 Prague}
\affiliation{Chiba University, Chiba 263-8522}
\affiliation{Chonnam National University, Kwangju 660-701}
\affiliation{University of Cincinnati, Cincinnati, Ohio 45221}
\affiliation{Deutsches Elektronen--Synchrotron, 22607 Hamburg}
\affiliation{Department of Physics, Fu Jen Catholic University, Taipei 24205}
\affiliation{Justus-Liebig-Universit\"at Gie\ss{}en, 35392 Gie\ss{}en}
\affiliation{Gifu University, Gifu 501-1193}
\affiliation{II. Physikalisches Institut, Georg-August-Universit\"at G\"ottingen, 37073 G\"ottingen}
\affiliation{The Graduate University for Advanced Studies, Hayama 240-0193}
\affiliation{Gyeongsang National University, Chinju 660-701}
\affiliation{Hanyang University, Seoul 133-791}
\affiliation{University of Hawaii, Honolulu, Hawaii 96822}
\affiliation{High Energy Accelerator Research Organization (KEK), Tsukuba 305-0801}
\affiliation{Hiroshima Institute of Technology, Hiroshima 731-5193}
\affiliation{IKERBASQUE, Basque Foundation for Science, 48011 Bilbao}
\affiliation{University of Illinois at Urbana-Champaign, Urbana, Illinois 61801}
\affiliation{Indian Institute of Technology Bhubaneswar, Satya Nagar 751007}
\affiliation{Indian Institute of Technology Guwahati, Assam 781039}
\affiliation{Indian Institute of Technology Madras, Chennai 600036}
\affiliation{Indiana University, Bloomington, Indiana 47408}
\affiliation{Institute of High Energy Physics, Chinese Academy of Sciences, Beijing 100049}
\affiliation{Institute of High Energy Physics, Vienna 1050}
\affiliation{Institute for High Energy Physics, Protvino 142281}
\affiliation{Institute of Mathematical Sciences, Chennai 600113}
\affiliation{INFN - Sezione di Torino, 10125 Torino}
\affiliation{Institute for Theoretical and Experimental Physics, Moscow 117218}
\affiliation{J. Stefan Institute, 1000 Ljubljana}
\affiliation{Kanagawa University, Yokohama 221-8686}
\affiliation{Institut f\"ur Experimentelle Kernphysik, Karlsruher Institut f\"ur Technologie, 76131 Karlsruhe}
\affiliation{Kavli Institute for the Physics and Mathematics of the Universe (WPI), University of Tokyo, Kashiwa 277-8583}
\affiliation{Department of Physics, Faculty of Science, King Abdulaziz University, Jeddah 21589}
\affiliation{Korea Institute of Science and Technology Information, Daejeon 305-806}
\affiliation{Korea University, Seoul 136-713}
\affiliation{Kyoto University, Kyoto 606-8502}
\affiliation{Kyungpook National University, Daegu 702-701}
\affiliation{\'Ecole Polytechnique F\'ed\'erale de Lausanne (EPFL), Lausanne 1015}
\affiliation{Faculty of Mathematics and Physics, University of Ljubljana, 1000 Ljubljana}
\affiliation{Luther College, Decorah, Iowa 52101}
\affiliation{University of Maribor, 2000 Maribor}
\affiliation{Max-Planck-Institut f\"ur Physik, 80805 M\"unchen}
\affiliation{School of Physics, University of Melbourne, Victoria 3010}
\affiliation{Moscow Physical Engineering Institute, Moscow 115409}
\affiliation{Moscow Institute of Physics and Technology, Moscow Region 141700}
\affiliation{Graduate School of Science, Nagoya University, Nagoya 464-8602}
\affiliation{Kobayashi-Maskawa Institute, Nagoya University, Nagoya 464-8602}
\affiliation{Nara University of Education, Nara 630-8528}
\affiliation{Nara Women's University, Nara 630-8506}
\affiliation{National Central University, Chung-li 32054}
\affiliation{National United University, Miao Li 36003}
\affiliation{Department of Physics, National Taiwan University, Taipei 10617}
\affiliation{H. Niewodniczanski Institute of Nuclear Physics, Krakow 31-342}
\affiliation{Nippon Dental University, Niigata 951-8580}
\affiliation{Niigata University, Niigata 950-2181}
\affiliation{University of Nova Gorica, 5000 Nova Gorica}
\affiliation{Osaka City University, Osaka 558-8585}
\affiliation{Osaka University, Osaka 565-0871}
\affiliation{Pacific Northwest National Laboratory, Richland, Washington 99352}
\affiliation{Panjab University, Chandigarh 160014}
\affiliation{Peking University, Beijing 100871}
\affiliation{University of Pittsburgh, Pittsburgh, Pennsylvania 15260}
\affiliation{Punjab Agricultural University, Ludhiana 141004}
\affiliation{Research Center for Electron Photon Science, Tohoku University, Sendai 980-8578}
\affiliation{Research Center for Nuclear Physics, Osaka University, Osaka 567-0047}
\affiliation{RIKEN BNL Research Center, Upton, New York 11973}
\affiliation{Saga University, Saga 840-8502}
\affiliation{University of Science and Technology of China, Hefei 230026}
\affiliation{Seoul National University, Seoul 151-742}
\affiliation{Shinshu University, Nagano 390-8621}
\affiliation{Soongsil University, Seoul 156-743}
\affiliation{Sungkyunkwan University, Suwon 440-746}
\affiliation{School of Physics, University of Sydney, NSW 2006}
\affiliation{Department of Physics, Faculty of Science, University of Tabuk, Tabuk 71451}
\affiliation{Tata Institute of Fundamental Research, Mumbai 400005}
\affiliation{Excellence Cluster Universe, Technische Universit\"at M\"unchen, 85748 Garching}
\affiliation{Toho University, Funabashi 274-8510}
\affiliation{Tohoku Gakuin University, Tagajo 985-8537}
\affiliation{Tohoku University, Sendai 980-8578}
\affiliation{Department of Physics, University of Tokyo, Tokyo 113-0033}
\affiliation{Tokyo Institute of Technology, Tokyo 152-8550}
\affiliation{Tokyo Metropolitan University, Tokyo 192-0397}
\affiliation{Tokyo University of Agriculture and Technology, Tokyo 184-8588}
\affiliation{University of Torino, 10124 Torino}
\affiliation{Toyama National College of Maritime Technology, Toyama 933-0293}
\affiliation{CNP, Virginia Polytechnic Institute and State University, Blacksburg, Virginia 24061}
\affiliation{Wayne State University, Detroit, Michigan 48202}
\affiliation{Yamagata University, Yamagata 990-8560}
\affiliation{Yonsei University, Seoul 120-749}
  \author{A.~Abdesselam}\affiliation{Department of Physics, Faculty of Science, University of Tabuk, Tabuk 71451} 
  \author{I.~Adachi}\affiliation{High Energy Accelerator Research Organization (KEK), Tsukuba 305-0801}\affiliation{The Graduate University for Advanced Studies, Hayama 240-0193} 
  \author{K.~Adamczyk}\affiliation{H. Niewodniczanski Institute of Nuclear Physics, Krakow 31-342} 
  \author{H.~Aihara}\affiliation{Department of Physics, University of Tokyo, Tokyo 113-0033} 
  \author{S.~Al~Said}\affiliation{Department of Physics, Faculty of Science, University of Tabuk, Tabuk 71451}\affiliation{Department of Physics, Faculty of Science, King Abdulaziz University, Jeddah 21589} 
  \author{K.~Arinstein}\affiliation{Budker Institute of Nuclear Physics SB RAS and Novosibirsk State University, Novosibirsk 630090} 
  \author{Y.~Arita}\affiliation{Graduate School of Science, Nagoya University, Nagoya 464-8602} 
  \author{D.~M.~Asner}\affiliation{Pacific Northwest National Laboratory, Richland, Washington 99352} 
  \author{T.~Aso}\affiliation{Toyama National College of Maritime Technology, Toyama 933-0293} 
  \author{V.~Aulchenko}\affiliation{Budker Institute of Nuclear Physics SB RAS and Novosibirsk State University, Novosibirsk 630090} 
% ...
\collaboration{The Belle Collaboration}
\noaffiliation

\begin{abstract}
We present a study of Michel parameters in leptonic $\tau$ decays 
using experimental information collected at the Belle detector. 
Michel parameters are extracted in the 
unbinned maximum likelihood fit of the 
$(\tau^{\mp}\to\ell^{\mp}\nu\nu,~\tau^{\pm}\to\pi^{\pm}\pi^0\nu)$
events in the full nine-dimensional phase space. We exploit the 
spin-spin correlation of tau leptons to extract $\xi_\rho\xi$ 
and $\xi_\rho\xi\delta$ in addition to the 
$\rho$ and $\eta$ parameters. 
\end{abstract}


\maketitle

\tighten

{\renewcommand{\thefootnote}{\fnsymbol{footnote}}}
\setcounter{footnote}{0}

\section{Introduction}

In the Standard Model (SM), the charged weak interaction is described by the 
exchange of a $W^{\pm}$ boson with a pure vector coupling to only
left-chirality fermions. Thus, in the low-energy four-fermion framework,
the Lorentz structure of the matrix element is predicted to be of the 
``V-A$\otimes$V-A'' type. Deviations from this behavior would indicate 
new physics and might be caused either by changes in the W-boson couplings or 
through interactions mediated by new gauge bosons. 
Leptonic decays such as 
$\tau^-\to\ell^-\bar{\nu_{\ell}}\nu_{\tau}~(\ell=e,\mu)$ 
(unless specified otherwise, charge-conjugated decays are implied throughout the paper)  
are the only ones in which the electroweak couplings can be probed without 
disturbance from the strong interaction. This makes them an ideal
system to study the Lorentz structure of the charged weak current. 

The most general, Lorentz invariant, derivative-free and 
lepton-number-conserving four-lepton point interaction 
matrix element for this decay can be written as \cite{Fetscher:1986uj}: 
\begin{equation}
{\cal M} = \frac{4G}{\sqrt{2}}\sum_{\begin{subarray}{c} N=S,V,T \\ 
i,j=L,R \end{subarray}} g_{ij}^{N} 
\biggl[\bar{u}_i(l^-)\Gamma^N v_n(\bar{\nu}_l)\biggr]
\biggl[\bar{u}_m(\nu_{\tau})\Gamma_N u_j(\tau^-)\biggr], 
\end{equation}
\begin{equation}  
\Gamma^{S}=1,~\Gamma^{V}=\gamma^{\mu},~\Gamma^{T}=\frac{1}{\sqrt{2}}\sigma^{\mu\nu}=\frac{i}{2\sqrt{2}}(\gamma^{\mu}\gamma^{\nu}-\gamma^{\nu}\gamma^{\mu})
\end{equation}
The $\Gamma_N$ matrices define the properties of the two currents
under a Lorentz transformation with $N=S,V,T$ for scalar, vector and
tensor interactions, respectively. The indices $i$ and $j$ label the 
right- or left-handedness (R, L) of the charged leptons. For a given $i$, $j$ and 
$N$, the handedness of the neutrinos ($n$, $m$) is fixed. Ten
non-trivial terms are characterized by ten complex coupling 
constants $g_{ij}^{N}$; those with $g^{T}_{RR}$ and $g^{T}_{LL}$ 
are identically zero. 
In the SM, the only non-zero coupling constant is $g^{V}_{LL}=1$. 
As the couplings can be complex, with arbitrary overall phase, there are 19 
independent parameters. The total strength of the weak interaction 
(charged weak current sector) is determined by the Fermi constant $G$, 
hence, the $g_{ij}^{N}$ are normalized as: 
\[ 3\biggl( |g^T_{LR}|^2+|g^T_{RL}|^2 \biggr)  
+\biggl( |g^V_{LL}|^2+|g^V_{LR}|^2+|g^V_{RL}|^2 +|g^V_{RR}|^2\biggr)+ \] 
\begin{equation}
+\frac{1}{4}\biggl(|g^S_{LL}|^2+|g^S_{LR}|^2+|g^S_{RL}|^2+|g^S_{RR}|^2\biggr)\equiv 1 
\end{equation} 
This constrains the coupling constants to be $|g^S_{ij}|\leq 2$, 
$|g^V_{ij}|\leq 1$ and $|g^T_{ij}|\leq 1/\sqrt{3}$. 
  
In the case where neutrinos are not detected and the spin of the
outgoing charged lepton is not determined, only four Michel 
parameters (MP) $\rho$, $\eta$, $\xi$ and $\delta$ are experimentally 
accessible. They are bilinear combinations 
of the $g_{ij}^{N}$ coupling constants \cite{Fetscher:1993ki}:
\begin{equation}
\rho = \frac{3}{4}-\frac{3}{4}\biggl(|g^V_{LR}|^2+|g^V_{RL}|^2+
2|g^T_{LR}|^2+2|g^T_{RL}|^2+\Re\bigl(g^S_{LR}g^{T*}_{LR}+
g^S_{RL}g^{T*}_{RL}\bigr)\biggr)
\end{equation}
\begin{equation}
\eta = \frac{1}{2}\Re\biggl(6g^V_{RL}g^{T*}_{LR}+
6g^V_{LR}g^{T*}_{RL}+g^S_{RR}g^{V*}_{LL}+g^S_{RL}g^{V*}_{LR}+
g^S_{LR}g^{V*}_{RL}+g^S_{LL}g^{V*}_{RR}\biggr) 
\end{equation}
\[ \xi = 4\Re(g^S_{LR}g^{T*}_{LR})-4\Re(g^S_{RL}g^{T*}_{RL})+
|g^V_{LL}|^2+3|g^V_{LR}|^2-3|g^V_{RL}|^2-|g^V_{RR}|^2 \] 
\begin{equation}
+5|g^T_{LR}|^2-5|g^T_{RL}|^2+\frac{1}{4}|g^S_{LL}|^2-\frac{1}{4}|g^S_{LR}|^2+
\frac{1}{4}|g^S_{RL}|^2-\frac{1}{4}|g^S_{RR}|^2 
\end{equation} 
\[ \xi\delta =\frac{3}{16}|g^S_{LL}|^2-\frac{3}{16}|g^S_{LR}|^2+
\frac{3}{16}|g^S_{RL}|^2-\frac{3}{16}|g^S_{RR}|^2-\frac{3}{4}|g^T_{LR}|^2+
\frac{3}{4}|g^T_{RL}|^2 \]
\begin{equation}
+\frac{3}{4}|g^V_{LL}|^2-\frac{3}{4}|g^V_{RR}|^2+ 
\frac{3}{4}\Re\bigl(g^S_{LR}g^{T*}_{LR}\bigr)-\frac{3}{4}\Re\bigl(g^S_{RL}g^{T*}_{RL}\bigr)
\end{equation}
and appear in the predicted energy spectrum of the charged lepton.

In the $\tau$ rest frame, neglecting radiative corrections, 
this spectrum is given by \cite{pdg}: 
\[ \frac{d\Gamma(\tau^{\mp})}{d\Omega dx} = \frac{4G^2 M_{\tau}E^4_{\rm max}}{(2\pi)^4}\sqrt{x^2-x^2_0}\biggl(x(1-x)+\frac{2}{9}\rho(4x^2-3x-x^2_0)+\eta x_0(1-x) \]
\[ \mp\frac{1}{3}P_{\tau}\cos\theta_{\ell}\xi\sqrt{x^2-x^2_0}\biggl[1-x+\frac{2}{3}\delta\bigl(4x-4+\sqrt{1-x^2_0}\bigr)\biggr]\biggr),\]
\begin{equation}
x=\frac{E_{\ell}}{E_{max}},~E_{max}=\frac{M_{\tau}}{2}(1+\frac{m^2_{\ell}}{M^2_{\tau}}),~x_0=\frac{m_{\ell}}{E_{max}}, 
\label{eqlepdec}
\end{equation} 
{\noindent where $P_{\tau}$ is $\tau$ polarization,} 
and $\theta_{\ell}$ is the angle between the $\tau$ spin and the
lepton momentum. In the SM, the ``V-A'' charged weak current is 
characterized by $\rho=3/4$, $\eta=0$, $\xi=1$ and $\delta=3/4$. 

\section{Method} 

Measurement of $\xi$ and $\delta$ requires knowledge of the  
$\tau$ spin direction. In experiments at $e^+e^-$ colliders 
with unpolarized $e^\pm$ beams, the average polarization of a single 
$\tau$ is zero. However, spin-spin correlations between 
the $\tau^+$ and $\tau^-$ produced in the reaction $e^+e^-\to\tau^+\tau^-$ 
can be exploited \cite{Tsai:1}. 
The main idea of our method is to consider events where both taus 
decay to selected final states. One (signal) tau 
decays leptonically ($\tau^-\to\ell^-\nu_\tau\bar{\nu}_\ell$,~$\ell=e,~\mu$) 
while the opposite tau, which decays via $\tau^+\to\pi^+\pi^0\bar{\nu}_\tau$, 
serves as a spin analyser. We choose the $\tau^+\to \pi^+\pi^0\bar{\nu}_\tau$ 
decay mode because it has the largest branching fraction as well as 
properly studied dynamics \cite{Fujikawa:1}. 
To write the total differential cross section for  
$(\tau^-\to\ell^-\nu_\tau\bar{\nu}_\ell,~\tau^+\to\pi^+\pi^0\bar{\nu}_\tau)$ 
(or, briefly, $\ell-\rho$) events, we follow the approach developed 
in Refs.~\cite{Fetscher:1,Tamai:1,Tamai:2}. 
The differential cross section of the $e^+e^-\to\tau^+(\vec{\zeta}^{*+})\tau^-(\vec{\zeta}^{*-})$ 
reaction in the center-of-mass system (c.m.s.) is given 
by their formula \cite{Tsai:1}:
\begin{equation}
\frac{d\sigma(\vec{\zeta}^{*-},\vec{\zeta}^{*+})}{d\Omega} = 
\frac{\alpha^2}{64E^2_{\tau}}\beta_\tau (D_0+D_{ij}\zeta^{*-}_i\zeta^{*+}_j),  
\end{equation} 
where $D_0 = 1+\cos^2{\theta}+\frac{1}{\gamma^2_{\tau}}\sin^2{\theta}$,  
$D_{ij}$ is the spin-spin correlation tensor, and 
$\vec{\zeta}^{*\mp}$ is the polarisation vector of the $\tau^{\mp}$
in the $\tau^{\mp}$ rest frame (unit vector along the $\tau^{\mp}$ spin direction). 
The asterisk denotes a parameter measured in the associated $\tau$ rest frame. 
The  differential decay width of the signal is written in the form
(with the total normalization constant $\kappa_\ell$ that is unimportant in this context): 
\[
\frac{d\Gamma(\tau^{\mp}(\vec{\zeta}^{*\mp})\to\ell^{\mp}\nu\nu)}{dx^*d\Omega^*_{\ell}}=\kappa_{\ell}(A(x^*)\mp\xi\vec{n}^*_{\ell}\vec{\zeta}^{*\mp}B(x^*)),
\] 
\begin{equation} 
A(x^*)=A_0(x^*)+\rho A_1(x^*)+\eta A_2(x^*),~B(x^*)=B_1(x^*)+\delta B_2(x^*), 
\label{eqlep}
\end{equation} 
where the form factors $A_0,~A_1,~A_2,~B_1$ and $B_2$ can be extracted from Eq.~\ref{eqlepdec}. 
The $\tau^{\pm}(\vec{\zeta'}^*)\to\rho^{\pm}(K^*)\nu(q^*)\to\pi^{\pm}(p^*_1)\pi^0(p^*_2)\nu(q^*)$ 
decay width reads (with the total normalization constant $\kappa_\rho$): 
\begin{equation}
\frac{d\Gamma(\tau^{\pm}\to\pi^{\pm}\pi^0\nu)}{dm^2_{\pi\pi}d\Omega^*_{\rho}d\tilde{\Omega}_{\pi}}=\kappa_{\rho}(A'\mp\vec{B'}\vec{\zeta'}^*) W(m^2_{\pi\pi}),
\end{equation}
\[ A'=2(q,Q)Q^*_0-Q^2q^*_0,~\vec{B'}=Q^2\vec{K}^*+2(q,Q)\vec{Q}^*,~Q^*=p^*_1-p^*_2,~K^*=p^*_1+p^*_2, \] 
\[ W(m^2_{\pi\pi})=|F_{\pi}(m^2_{\pi\pi})|^2\frac{p^*_\rho(m^2_{\pi\pi})\tilde{p}_{\pi}(m^2_{\pi\pi})}{M_{\tau}m_{\pi\pi}},~m^2_{\pi\pi}=K^{*2},~p^*_\rho=\frac{M_{\tau}}{2}\biggl(1-\frac{m^2_{\pi\pi}}{M^2_{\tau}}\biggr),\] 
\begin{equation} 
\tilde{p}_{\pi}=\frac{\sqrt{(m^2_{\pi\pi}-(m_{\pi}+m_{\pi^0})^2)(m^2_{\pi\pi}-(m_{\pi}-m_{\pi^0})^2)}}{2m_{\pi\pi}}, 
\end{equation} 
where $p^*_\rho$ and $\Omega^*_\rho$ are the momentum and solid angle
of the $\rho$ meson in the $\tau$ rest frame; 
$\tilde{p}_\pi$ and $\tilde{\Omega}_\pi$ are the momentum and solid
angle of the charged pion in the $\rho$ rest frame; and 
$F_{\pi}(m^2_{\pi\pi})$ is the pion form factor taken from Ref.~\cite{Fujikawa:1}. 
The total differential cross section for $\ell-\rho$ events is: 
\begin{equation}
\frac{d\sigma(\ell^{\mp},\rho^{\pm})}{dE^{*}_{\ell}d\Omega^{*}_{\ell}d\Omega^{*}_{\rho}dm^2_{\pi\pi}d\tilde{\Omega}_{\pi}d\Omega_{\tau}}=\kappa_\ell\kappa_\rho\frac{\alpha^2\beta_{\tau}}{64E^2_{\tau}}\bigl(D_0A'A+\xi_\rho\xi D_{ij}n^*_{\ell i}B'_jB\bigr)W(m^2_{\pi\pi}) 
\label{eqdiff}
\end{equation}
Experimentally, we measure particle parameters in the c.m.s.; hence, 
the visible differential cross section is given by \cite{Tamai:1}: 
\begin{equation}
\frac{d\sigma(\ell^{\mp},\rho^{\pm})}{dp_{\ell}d\Omega_{\ell}dp_{\rho}d\Omega_{\rho}dm^2_{\pi\pi}d\tilde{\Omega}_{\pi}}=\int\limits^{\Phi_2}_{\Phi_1}\frac{d\sigma(\ell^{\mp},\rho^{\pm})}{dE^{*}_{\ell}d\Omega^{*}_{\ell}d\Omega^{*}_{\rho}dm^2_{\pi\pi}d\tilde{\Omega}_{\pi}d\Omega_{\tau}}\biggl|\frac{\partial
  (E^*_{\ell},\Omega^*_{\ell},\Omega^*_{\rho},\Omega_{\tau})}{\partial
  (p_{\ell},\Omega_{\ell},p_{\rho},\Omega_{\rho},\Phi_{\tau})}\biggr|d\Phi_{\tau}, 
\label{eqdifint}
\end{equation} 
where the integration is performed over the unknown $\tau$ direction, which 
is constrained by the $(\Phi_1,\Phi_2)$ arc. Both $\Phi_1$ and $\Phi_2$ 
are calculated using parameters measured in the experiment. 
The differential cross section is used to construct the probability density 
function (p.d.f.) for the measurement vector 
$\vec{z}=(p_{\ell},~\cos{\theta_{\ell}},~\phi_{\ell},~p_{\rho},~\cos{\theta_{\rho}},~
\phi_{\rho},~m_{\pi\pi},~\cos{\tilde{\theta}_{\pi}},~\tilde{\phi}_{\pi})$:   
\[ 
{\cal P}(\vec{z})=\frac{{\cal F}(\vec{z})}{\int{\cal F}(\vec{z})d\vec{z}},~{\cal F}(\vec{z})=\frac{d\sigma(\ell^{\mp},\rho^{\pm})}{dp_{\ell}d\Omega_{\ell}dp_{\rho}d\Omega_{\rho}dm^2_{\pi\pi}d\tilde{\Omega}_{\pi}}={\cal F}_0 + {\cal F}_1\rho + {\cal F}_2\eta + {\cal F}_3\xi_\rho\xi + {\cal F}_4\xi_\rho\xi\delta, 
\]
\[ 
{\cal N}=\int{\cal F}(\vec{z})d\vec{z}={\cal N}_0 + {\cal N}_1\rho +
{\cal N}_2\eta + {\cal N}_3\xi_\rho\xi + {\cal
  N}_4\xi_\rho\xi\delta,~{\cal N}_i=\int{\cal
  F}_i(\vec{z})d\vec{z},~i=0\hbox{...}4, 
\]
\begin{equation} 
{\cal P}(\vec{z}) = \frac{{\cal F}_0(\vec{z}) + 
{\cal F}_1(\vec{z})\rho + 
{\cal F}_2(\vec{z})\eta + 
{\cal F}_3(\vec{z})\xi_\rho\xi + 
{\cal F}_4(\vec{z})\xi_\rho\xi\delta}{
{\cal N}_0 + 
{\cal N}_1\rho +
{\cal N}_2\eta + 
{\cal N}_3\xi_\rho\xi + 
{\cal N}_4\xi_\rho\xi\delta},  
\end{equation} 
where the form factors ${\cal F}_i$ are calculated for each event  
and the five normalisation constants ${\cal N}_i$ are evaluated using 
a Monte Carlo (MC) simulated sample. 
There are several corrections that must be incorporated in the 
procedure to take into account the real experimental
situation. Physics corrections include electroweak higher-order 
corrections to the $e^+ e^-\to\tau^+\tau^-$ cross section 
\cite{Arbuzov:1997pj,Arbuzov:2004,Arbuzov:2005pt,Kuraev:1985hb,Berends:1982dy,Jadach:1985ac,Jadach:1984iy}, 
the effect of the radiative leptonic decay 
$\tau^-\to\ell^-\bar{\nu}_{\ell}\nu_{\tau}\gamma$ \cite{Arbuzov:1,Arbuzov:2,Arbuzov:3}, 
and the effect of the radiative hadronic decay 
$\tau^-\to\pi^-\pi^0\nu_{\tau}\gamma$ \cite{Flores:1,Flores:2}. 
Apparatus corrections include the effect of the finite detection 
efficiency and resolution, the effect of the external bremsstrahlung 
for $e-\rho$ events, and the $e^\pm$ beam energy spread. 
 
The method described is used for a precise measurement of Michel 
parameters in $\ell-\rho$ events. This analysis is based on a 
$485~{\rm fb}^{-1}$ data sample that contains 446 
$\times 10^6\ \tau^+\tau^-$ pairs, 
collected with the Belle detector at the KEKB energy-asymmetric 
$e^+e^-$ (3.5 on 8~GeV) collider~\cite{kekb}
operating at the $\Upsilon(4S)$ resonance.

\section{The Belle detector}

The Belle detector is a large-solid-angle magnetic spectrometer that
consists of a silicon vertex detector (SVD),
a 50-layer central drift chamber (CDC), an array of
aerogel threshold Cherenkov counters (ACC), 
a barrel-like arrangement of time-of-flight
scintillation counters (TOF), and an electromagnetic calorimeter (ECL)
comprised of CsI(Tl) crystals located inside 
a superconducting solenoid coil that provides a 1.5~T
magnetic field.  An iron flux-return located outside 
the coil is instrumented to detect $K_L^0$ mesons and to identify
muons (KLM).  
Two inner detector configurations are used in this analysis. A beampipe 
with a radius of 2.0 cm and a 3-layer silicon vertex detector 
are used for the first sample
of $124\times 10^6$~$\tau^+\tau^-$ pairs, while a 1.5 cm beampipe, a 4-layer
silicon detector and a small-cell inner drift chamber are used to record  
the remaining $322\times 10^6$~$\tau^+\tau^-$ pairs~\cite{natk}.  
The detector is described in detail elsewhere~\cite{bel}.

\section{Selection of $\ell-\rho$ events, background}
\label{secsel}

This analysis is based on events with one $\tau$ decaying 
to leptons $\tau^-\to\ell^-\bar{\nu}_\ell\nu_\tau$ and the 
other decaying via the hadronic channel $\tau^+\to\pi^+\pi^0\bar{\nu}_\tau$. 

The selection process, which is designed to suppress background
while retaining a high efficiency for the decays under study,
proceeds in two stages.  

1) The first-stage criteria suppress beam background
to a negligible level and reject most of the background from  
other physical processes: 
\begin{itemize}
 \item There should be exactly two tracks extrapolated to the interaction 
       point within $\pm 0.5$~cm in the transverse direction 
       and $\pm 2.5$~cm along the beam and having a transverse 
       momentum in the c.m.s. $|\vec{P}|_{\rm{\perp}}^{\rm{CM}}>0.1$~GeV/$c$ 
       and a net charge of zero.
 \item The sum of the track absolute momenta in the c.m.s. 
       must satisfy $P^{\rm{CM}}<9$~GeV/$c$. 
 \item The maximum value of the transverse momentum for all tracks in the
       laboratory frame should satisfy $|\vec{P}|_{\rm{\perp}}^{\rm{LAB}}>0.5$~GeV/$c$.
 \item The maximum opening angle $\psi$ between tracks should exceed $20^\circ$.    
 \item The number of photons $N_\gamma$ with c.m.s. energy
       $E_{\rm{\gamma}}^{\rm{CM}}>80$~MeV should be five or fewer. 
 \item The total ECL energy deposition in the laboratory frame should satisfy  
       $\sum\limits_{i=1}^{N_{\rm clusters}} E_{i}^{\rm{LAB}}\rm{(ECL)} < 9$~GeV.
 \item The total energy of additional photons in the laboratory 
       frame should be $\sum E_{\rm{rest}\gamma}^{\rm{LAB}}<0.2$~GeV. 
 \item The missing mass should lie in the range 
       $1$~GeV/$c^2\leq M_{\rm{miss}} \leq 7$~GeV/$c^2$. 
 \item The polar angle of the missing momentum in the c.m.s. should 
       satisfy $30^\circ \leq \theta_{\rm{miss}}^{\rm{CM}}\leq 150^\circ$. 
\end{itemize}

The last two criteria are especially effective in suppressing background from  
radiative Bhabha, $\mu^+\mu^-$ and two-photon processes. 

2) In the second stage, the $\ell^\mp-\rho^\pm$
($\ell=e,\mu$,~$\rho^\pm\to\pi^\pm\pi^0$) samples are selected from
the remaining events. 

To select electrons, a likelihood ratio cut 
${\cal P}_{e}={\cal L}_{e}/({\cal L}_{e}+{\cal L}_{x})>0.8$ is applied,
where the electron likelihood function ${\cal L}_{e}$ and the non-electron
function ${\cal L}_{x}$ include information on the specific ionization 
($dE/dx$) measurement by the CDC, the ratio of the cluster energy in 
the ECL to the track momentum measured in the CDC, the transverse 
ECL shower shape and the light yield in the ACC \cite{eid}.
The efficiency of this cut for electrons is $93.1\%$.

To select muons, the likelihood ratio cut
${\cal P}_\mu={\cal L}_\mu /({\cal L}_\mu +{\cal L}_\pi +{\cal L}_K)>0.8$ 
is applied. It has an $88.0\%$ efficiency for muons. 
Each of the muon(${\cal L}_\mu$), pion(${\cal L}_\pi$) and 
kaon(${\cal L}_K$) likelihood functions is evaluated from 
two variables: the difference between the range calculated 
from the momentum of the particle and the range measured by KLM 
and the $\chi^2$ of the KLM hits with respect to the extrapolated 
track \cite{muid}. 

To separate pions from kaons, we determine for each track the 
pion (${\cal L}_\pi$) and kaon (${\cal L}_K$) likelihoods from 
the ACC response, the $dE/dx$ measurement in the CDC and the TOF
flight-time measurement, and form a likelihood ratio 
${\cal P}_{K/\pi}={\cal L}_K /({\cal L}_\pi +{\cal L}_K)$
to separate pions and kaons. For pions, we require
$\mathcal{P}_{K/\pi}<0.6$, which provides a pion identification 
efficiency of about $93\%$ while keeping the pion fake rate at 
the $6\%$ level.

Finally, we select events with only one lepton $\ell^\mp$ 
($\ell = e,\mu$), one charged pion $\pi^\pm$ and one $\pi^0$
candidate. A $\pi^0$ meson is reconstructed from a pair of gammas 
with an energy in the laboratory frame of $E^{\rm LAB}_{\gamma}>80$~MeV and   
the $\gamma\gamma$ invariant mass in the range 
$115$~MeV/$c^2<M_{\gamma\gamma}<150$~MeV/$c^2$. The absolute value of 
the $\pi^0$ momentum in the c.m.s. must satisfy $P^{\rm CMS}_{\pi^0}>0.3$~GeV/$c$. 
The invariant mass of the $\pi^\pm\pi^0$ system must lie in the range 
$0.3$~GeV/$c^2<M_{\pi^\pm\pi^0}<1.8$~GeV/$c^2$.  
The opening angles between a lepton and charged pion and between a 
lepton and $\pi^0$ should exceed $90^\circ$. 
To avoid the uncertainty due to the simulation of low 
energy fake ECL clusters, we allow additional photons 
in an event with the total energy in the laboratory 
frame of $E_{{\rm rest}\gamma}^{\rm LAB}<0.2$~GeV. 

To evaluate the background and calculate efficiencies, a 
Monte Carlo sample of $2.87 \times 10^9~\tau^+\tau^-$ pairs 
is produced with the KKMC/TAUOLA generators \cite{Jadach:1,tola}. The 
detector response is simulated by a GEANT3-based program \cite{geant}. 

The detection efficiencies for signal events are 
$\varepsilon_{\rm det}(e-\rho)=(11.53\pm 0.01)\%$ and  
$\varepsilon_{\rm det}(\mu-\rho)=(12.43\pm 0.01)\%$. 
It is found that the dominant background comes from other $\tau$
decays; a contribution from non-$\tau\tau$ processes 
is very small -- less than $0.1\%$. 
The dominant background arises from 
$(\tau^-\to\ell^-\bar{\nu}_\ell\nu_\tau,~\tau^+\to\pi^+\pi^0\pi^0\bar{\nu}_\tau)$ 
(or, briefly, $\ell-3\pi$) events, where the second $\pi^0$ is lost. 
Its contribution is $\lambda_{3\pi}=10.0$\% for the $e-\rho$ and 
$\lambda_{3\pi}=8.1$\% for the $\mu-\rho$ events. 
For the $\mu-\rho$ events, an additional background at the level 
of $\lambda_\pi=1.4$\% originates from $(\tau^-\to\pi^-\nu_\tau,~\tau^+\to\pi^+\pi^0\bar{\nu}_\tau)$ 
(or, briefly, $\pi-\rho$) events, where a pion is misidentified as a muon. 
The remaining background comes from other $\tau$ decays;  
its contribution is $\lambda_{\rm other}=2.0$\% for $e-\rho$ events 
and $\lambda_{\rm other}=2.5$\% for $\mu-\rho$ events. 

The main background processes, $\ell-3\pi$ and $\pi-\rho$, are included 
in the p.d.f. analytically while the remaining background is taken into 
account using the MC-based approach \cite{Schmidt:1}. 
The total p.d.f. is written as: 
\[ {\cal P}(\vec{z}) =
\frac{\varepsilon(\vec{z})}{\varepsilon}\biggl((1-\lambda_{3\pi}-\lambda_{\pi}-\lambda_{\rm other})\frac{S(\vec{z})}{\int\frac{\varepsilon(\vec{z})}{\varepsilon}S(\vec{z})d\vec{z}}+\lambda_{3\pi}\frac{\tilde{B}_{3\pi}(\vec{z})}{\int\frac{\varepsilon(\vec{z})}{\varepsilon}\tilde{B}_{3\pi}(\vec{z})d\vec{z}}+\]
\begin{equation}
+\lambda_{\pi}\frac{\tilde{B}_{\pi}(\vec{z})}{\int\frac{\varepsilon(\vec{z})}{\varepsilon}\tilde{B}_{\pi}(\vec{z})d\vec{z}}+
\lambda_{\rm other}\frac{B^{MC}_{\rm other}(\vec{z})}{\int\frac{\varepsilon(\vec{z})}{\varepsilon}
B^{MC}_{\rm other}(\vec{z})d\vec{z}}\biggr), 
\label{allbgs}
\end{equation}
where
$S(\vec{z})$,~$\tilde{B}_{3\pi}(\vec{z})$ and $\tilde{B}_{\pi}(\vec{z})$
are the cross sections for the $\ell-\rho$, $\ell-3\pi$ and $\pi-\rho$
events, respectively; $\varepsilon(\vec{z})$ is the detection
efficiency for signal events in the nine-dimensional phase space; 
and $\varepsilon=\int\varepsilon(\vec{z})S(\vec{z})d\vec{z}/\int
S(\vec{z})d\vec{z}$ is the average signal detection efficiency. 
   
\section{Analysis of experimental data}

\begin{figure}[htbp] 
\includegraphics[width=0.48\textwidth]{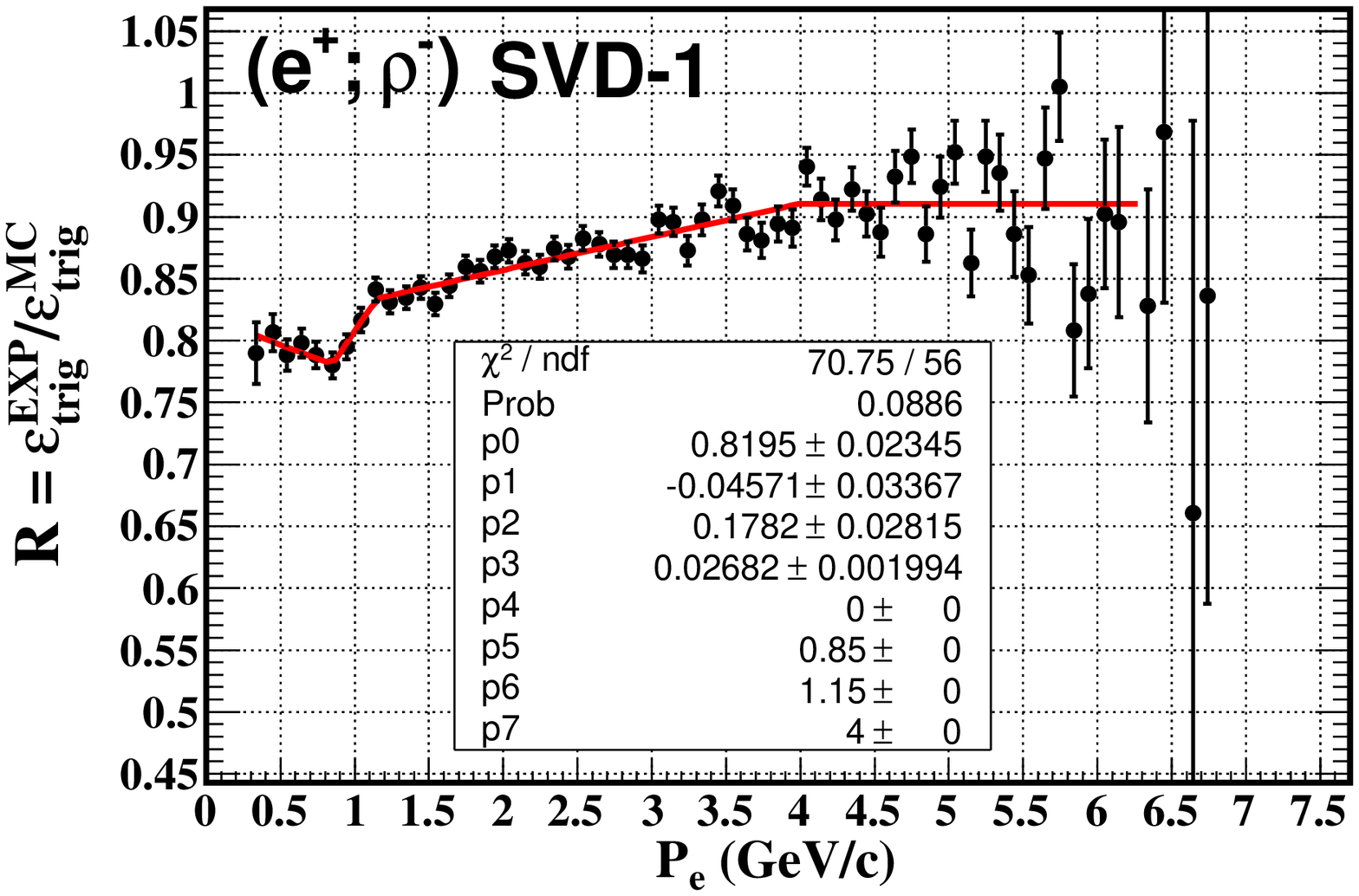} 
\includegraphics[width=0.48\textwidth]{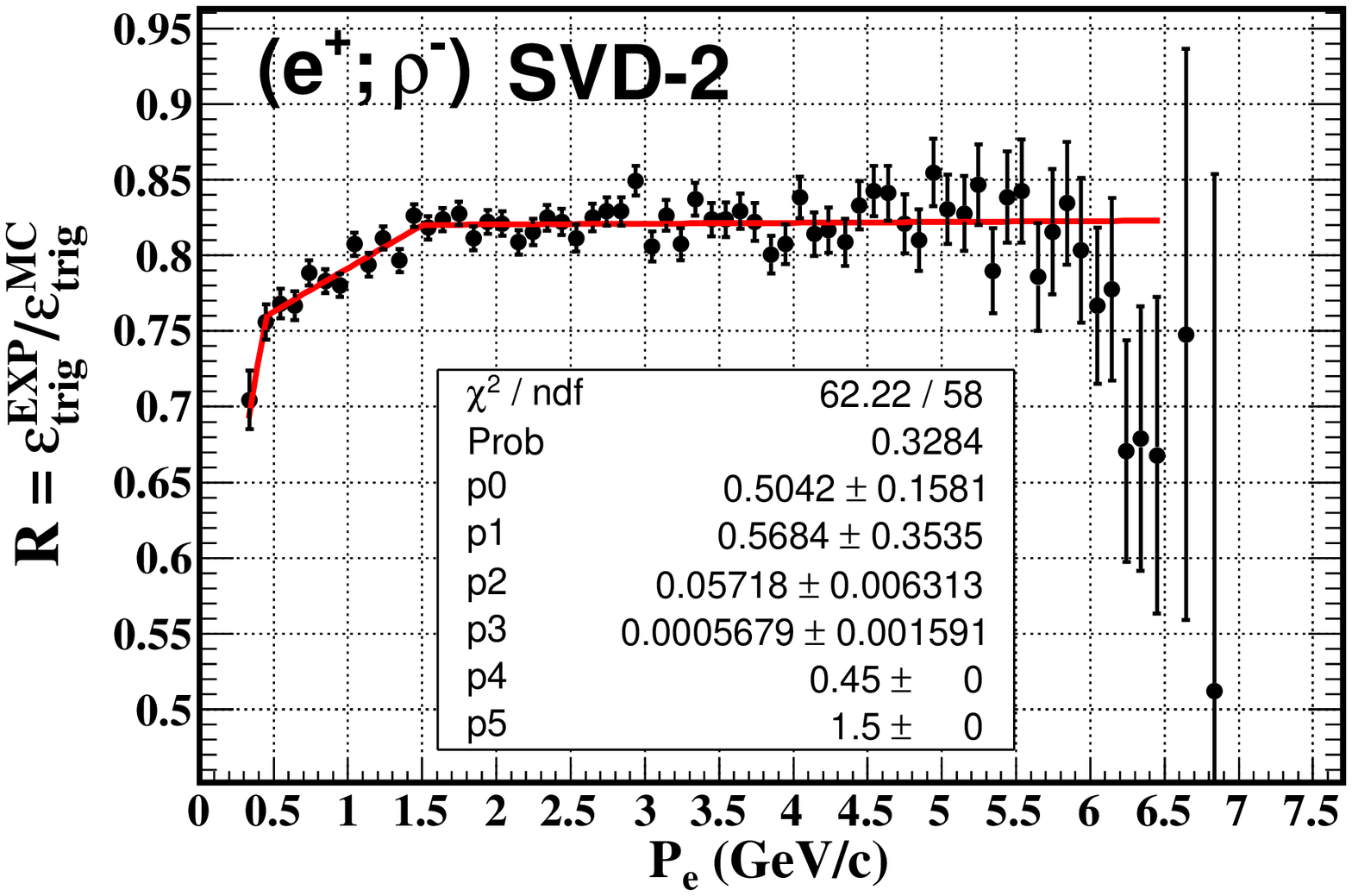} \\
\caption{Trigger efficiency correction for SVD-1 (left) and SVD-2 (right) 
$(e^+,~\rho^-)$ data samples as a function of electron momentum in the 
laboratory frame - black points with errors. It is fitted by an empirical 
function shown by the red solid line.}\label{trgeff} 
\end{figure}

After all selections,  about 5.5 million events in all four configurations 
($(e^+,~\pi^-\pi^0)$, $(e^-,~\pi^+\pi^0)$, $(\mu^+,~\pi^-\pi^0)$, $(\mu^-,~\pi^+\pi^0)$) 
are selected for the fit.  
\begin{figure}[htbp]
\includegraphics[width=0.48\textwidth]{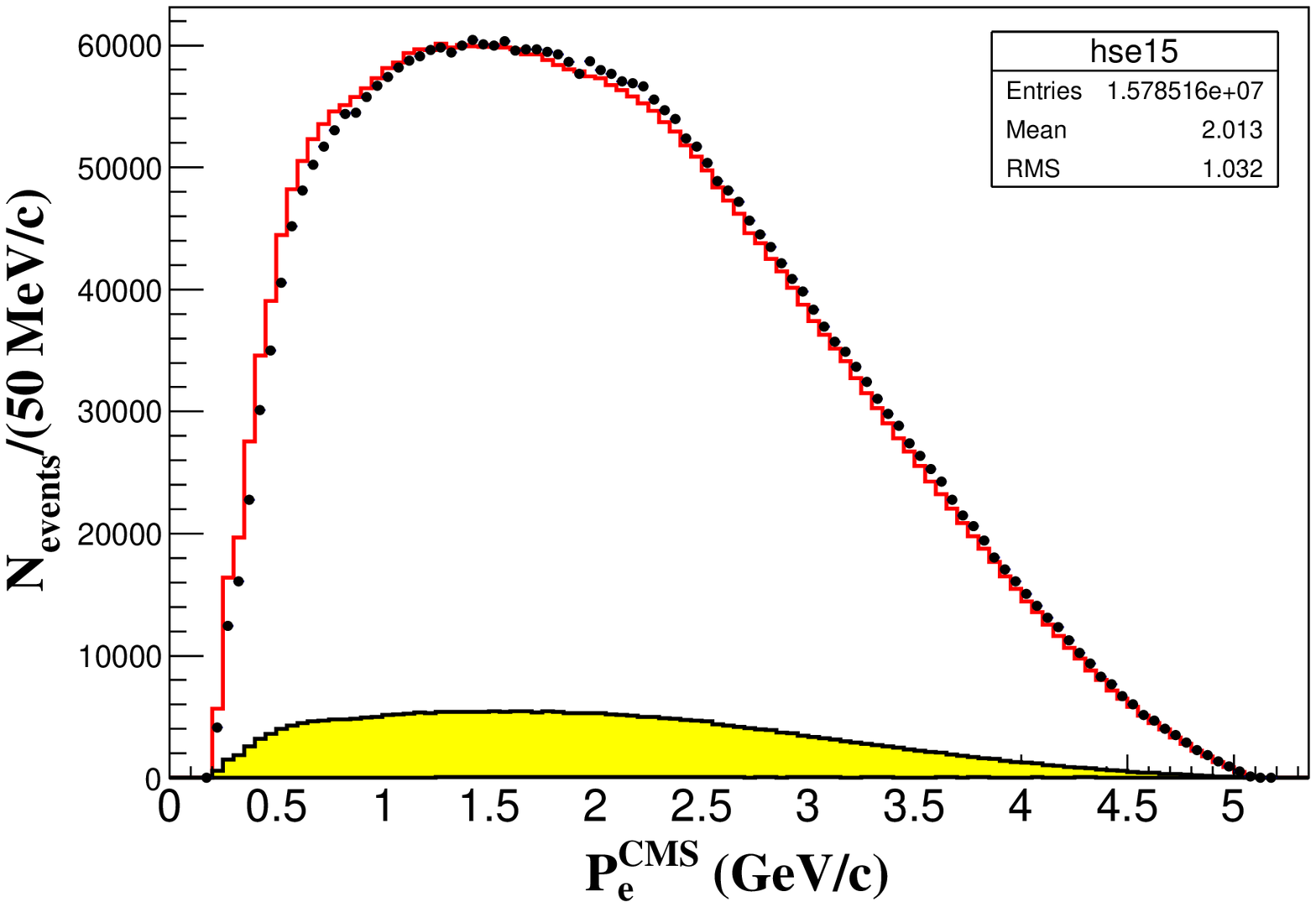} 
\includegraphics[width=0.48\textwidth]{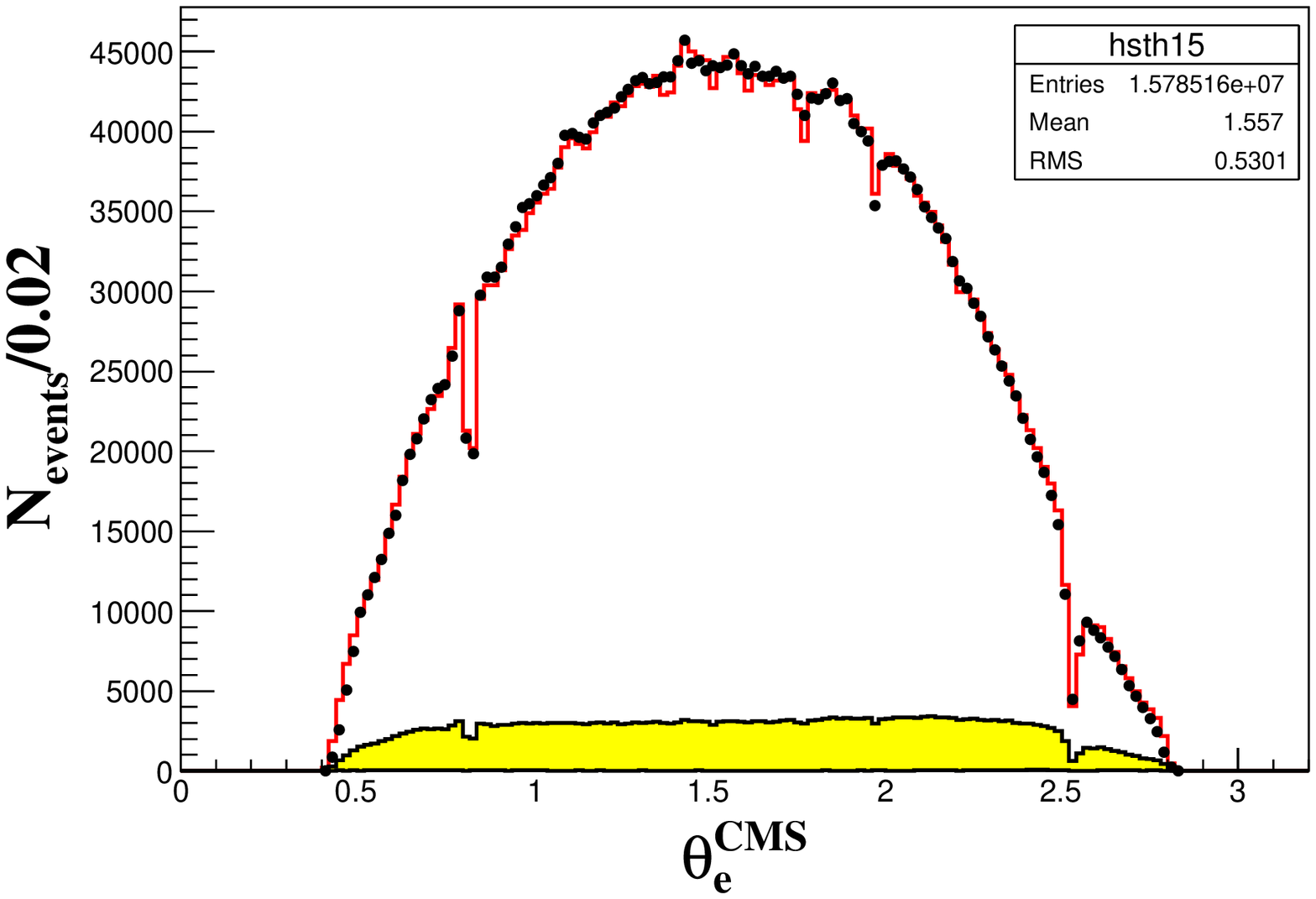} \\ 
\includegraphics[width=0.48\textwidth]{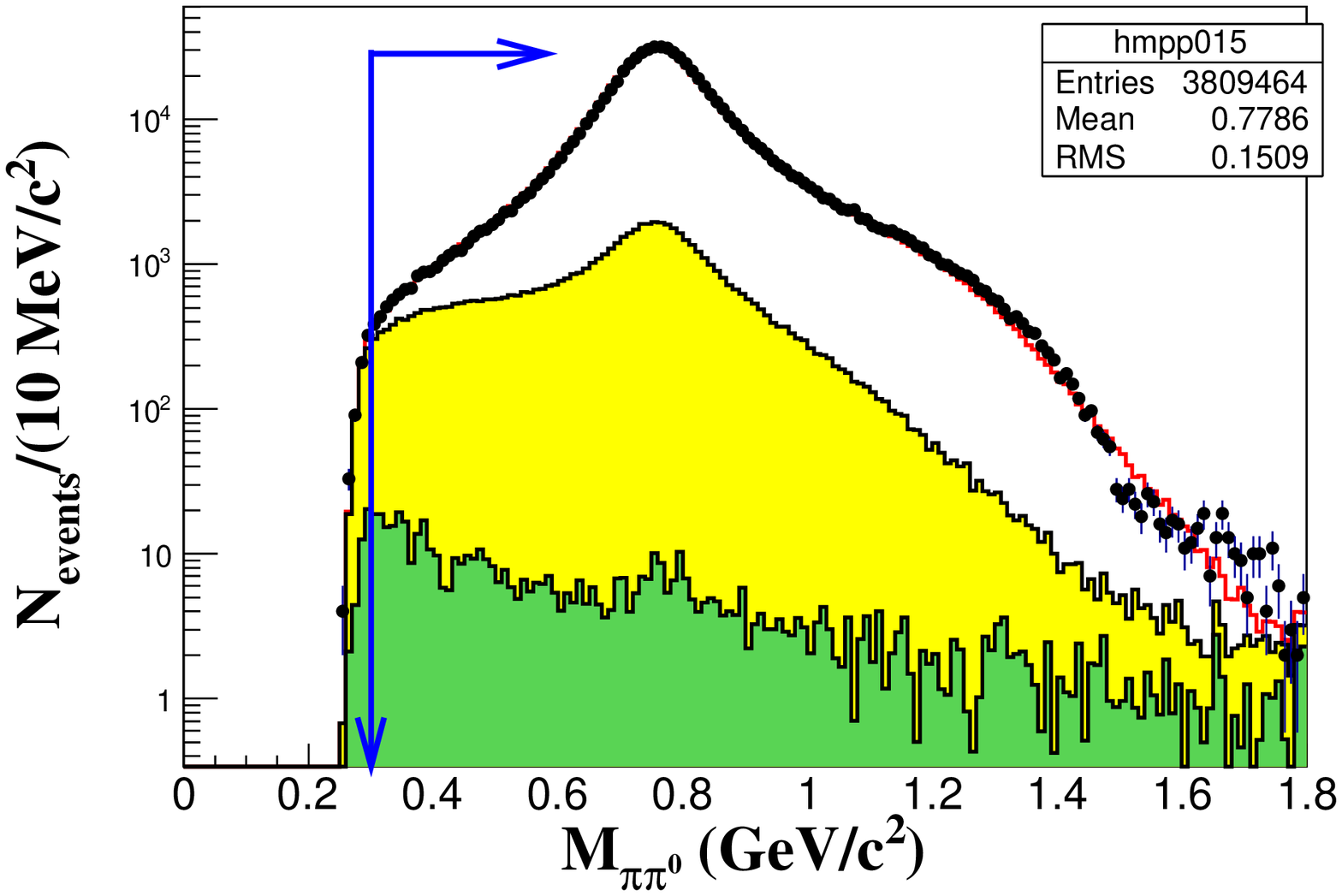} 
\includegraphics[width=0.48\textwidth]{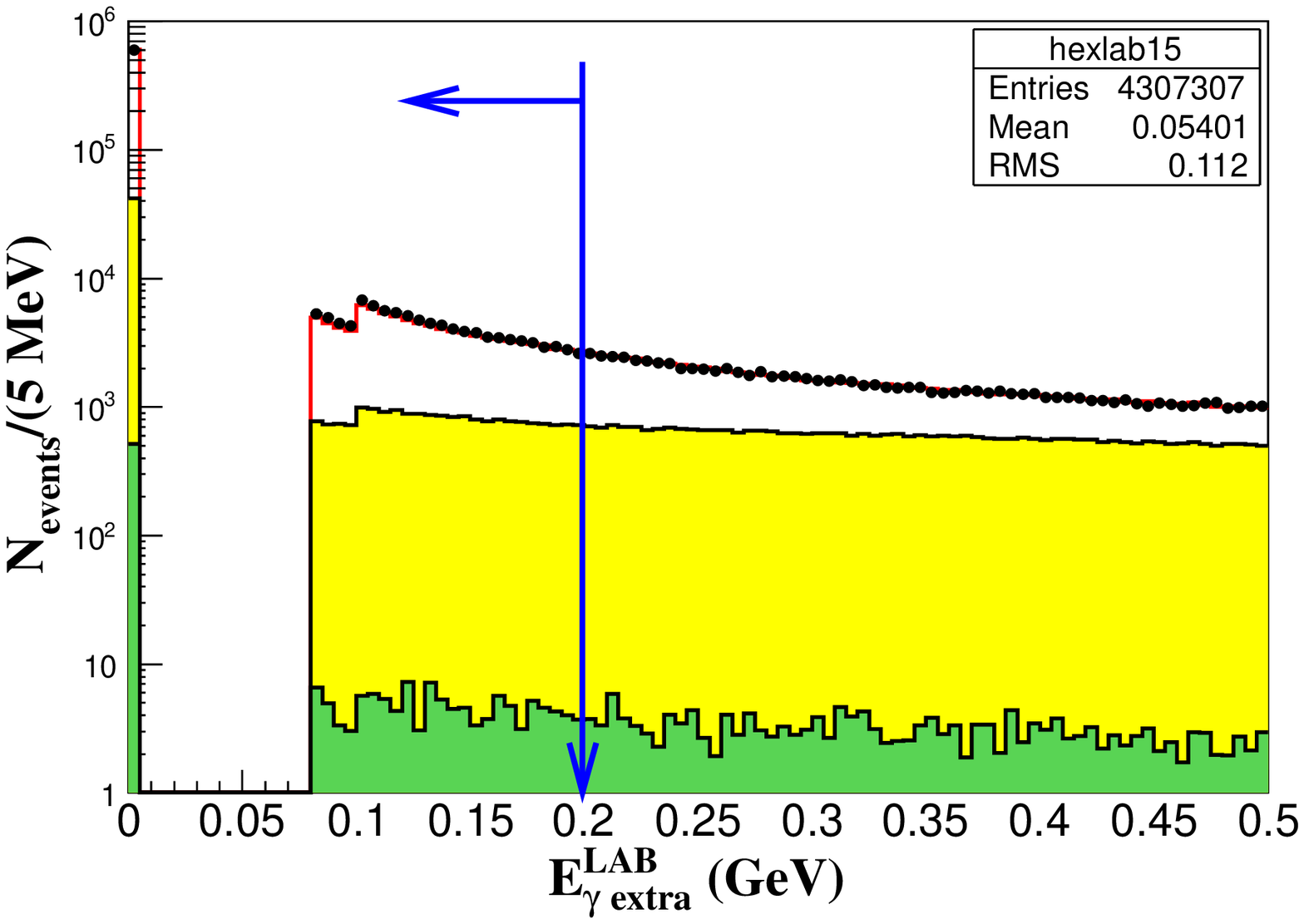} \\ 
\caption{Distributions of the selected $(e^+,~\pi^-\pi^0)$ events: 
$e^+$ momentum (upper left) and polar angle (upper right) in the c.m.s., 
$\pi^-\pi^0$ invariant mass (lower left), extra gamma energy in 
laboratory frame (lower right). Open histograms - signal MC simulation, 
yellow shaded histograms - the main background components from the 
$(e^+,~\pi^-\pi^0\pi^0)$ events, green shaded histograms - the
remaining background, points with errors - experimental data. 
Blue arrows show applied selections. MC and experimental histograms 
are normalized to the same number of events.}\label{selpic} 
\end{figure}
Figure~\ref{selpic} shows the distributions of selected kinematical 
parameters for $(e^+,~\pi^-\pi^0)$ events. Clearly, 
the experimental electron momentum spectrum is shifted markedly to 
higher momenta in comparison with the MC one. This is an artifact of the 
strong nonuniformity of the experimental trigger efficiency, which 
is not properly simulated. A special procedure has been 
developed to evaluate the trigger efficiency corrections, 
$\epsilon^{\rm TRG}_{\rm corr}=\varepsilon^{\rm TRG}_{\rm EXP}/\varepsilon^{\rm TRG}_{\rm MC}$;  
see Fig.~\ref{trgeff}. 
The trigger efficiency correction as well as the lepton identification efficiency
correction, $\epsilon^{\ell\rm ID}_{\rm corr}$, are incorporated 
in the fitter by modifying the detection efficiency in Eq.~\ref{allbgs}: 
$\varepsilon(\vec{z})\to \varepsilon(\vec{z})\epsilon^{\rm TRG}_{\rm 
corr}(p^{\rm LAB}_\ell)\epsilon^{\ell\rm ID}_{\rm corr}(p^{\rm LAB}_\ell)$. 

\begin{figure}[htbp]
\includegraphics[width=0.48\textwidth]{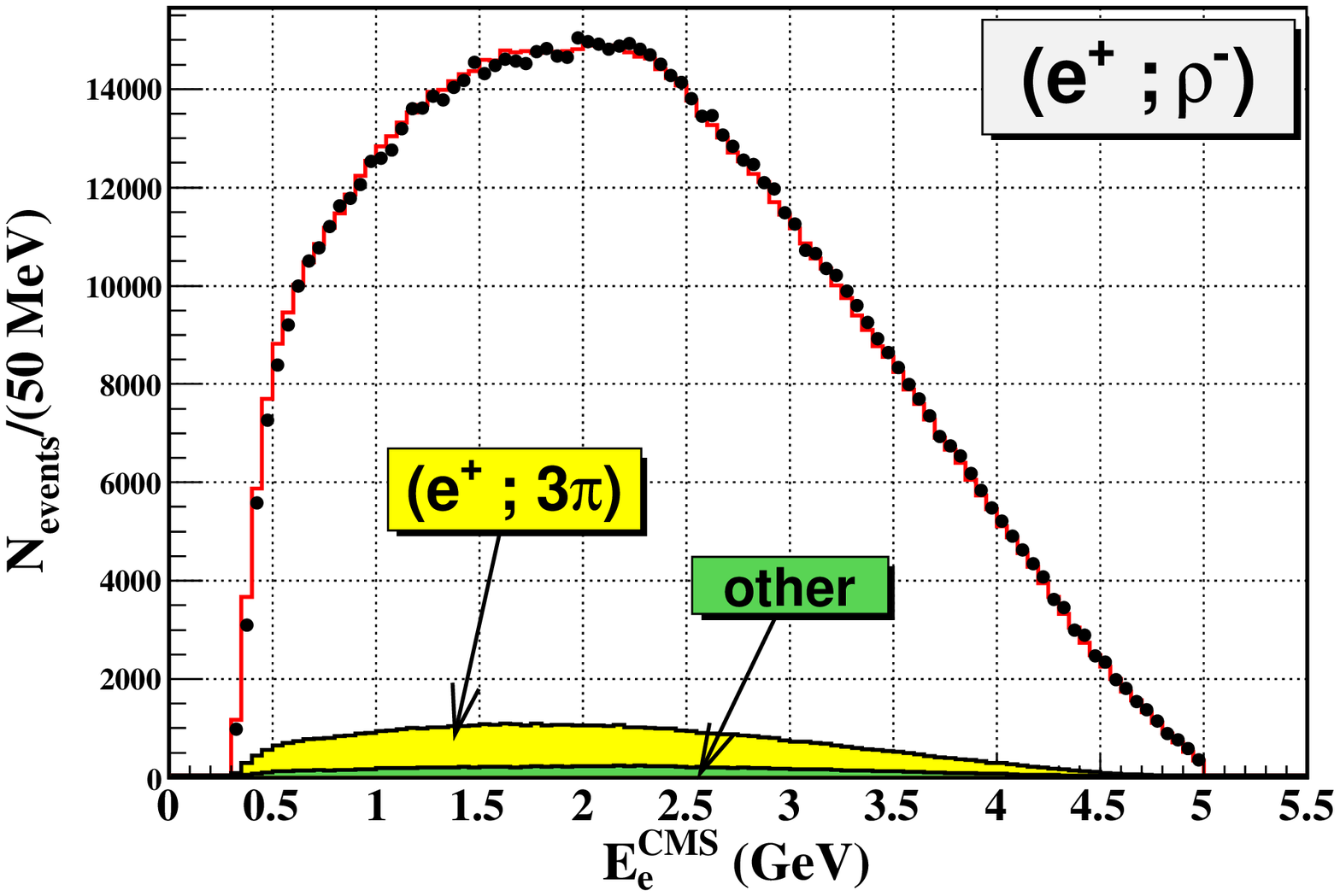} 
\includegraphics[width=0.48\textwidth]{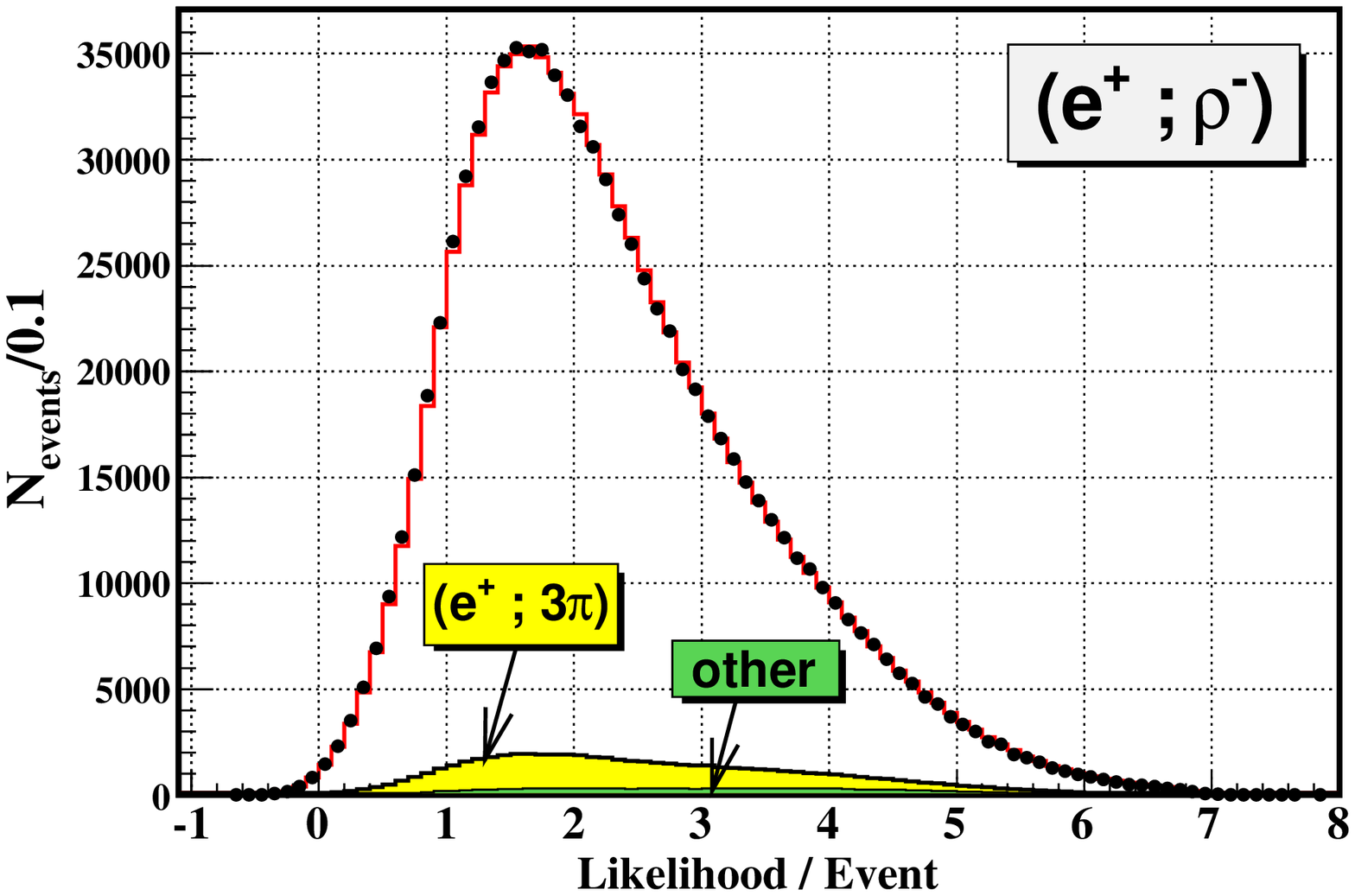}
\includegraphics[width=0.48\textwidth]{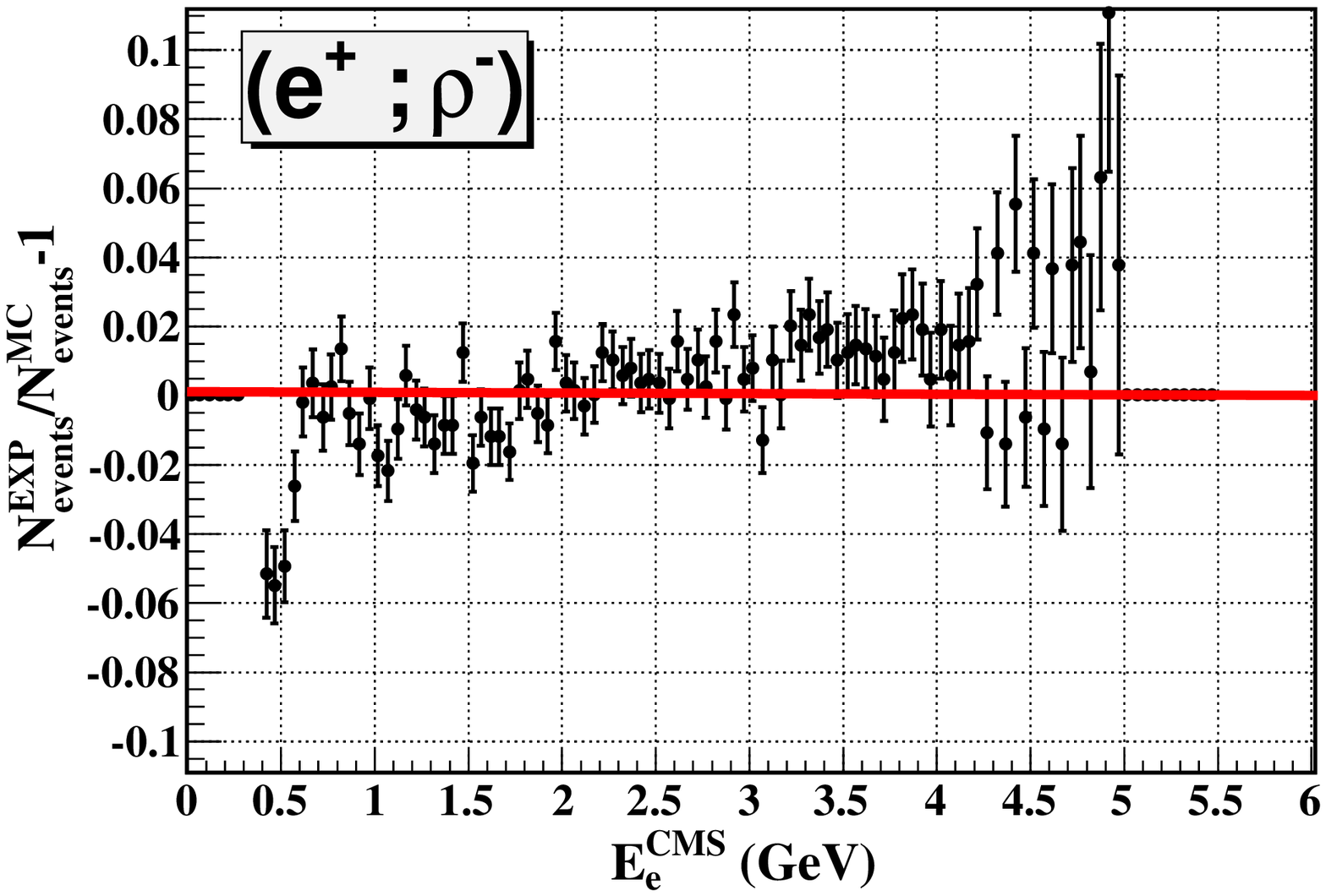} 
\includegraphics[width=0.48\textwidth]{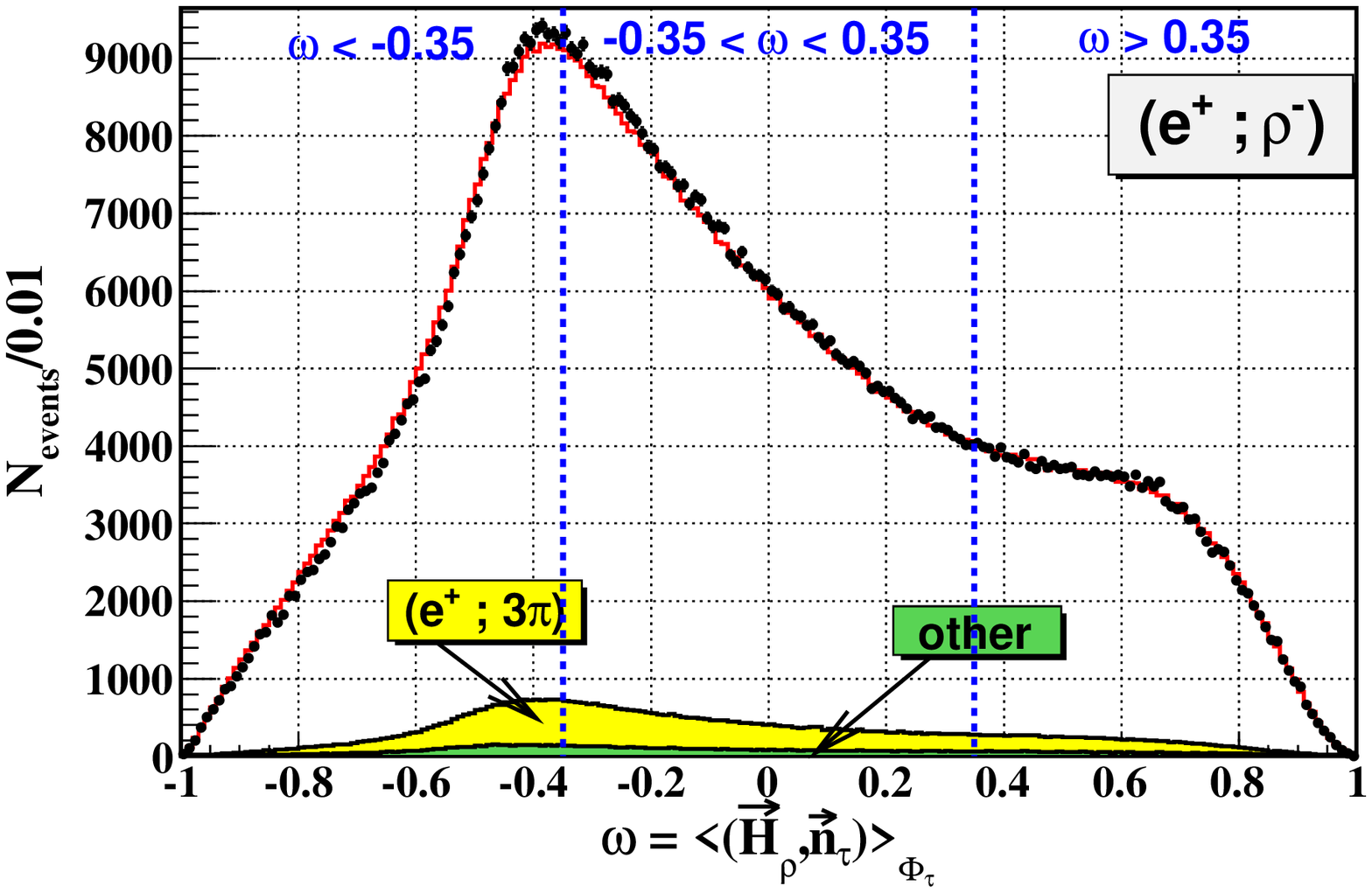} \\
\caption{Result of the fit of $(e^+,~\rho^-)$ experimental events. 
$e^+$ energy spectrum in CMS (upper left), relative difference between 
experimental energy spectrum and fit result (lower left), 
likelihood per event (upper right) and distribution of $\tau$ helicity 
sensitive variable $\omega$ (lower right). Points with errors show 
experimental data, histogram - result of the fit. Open histograms 
show signal events, shaded histograms - background contributions.
}\label{fitres} 
\end{figure}

\begin{figure}[htbp] 
\begin{tabular}{@{}c@{~}c@{~}c@{}}
\includegraphics[width=0.33\textwidth]{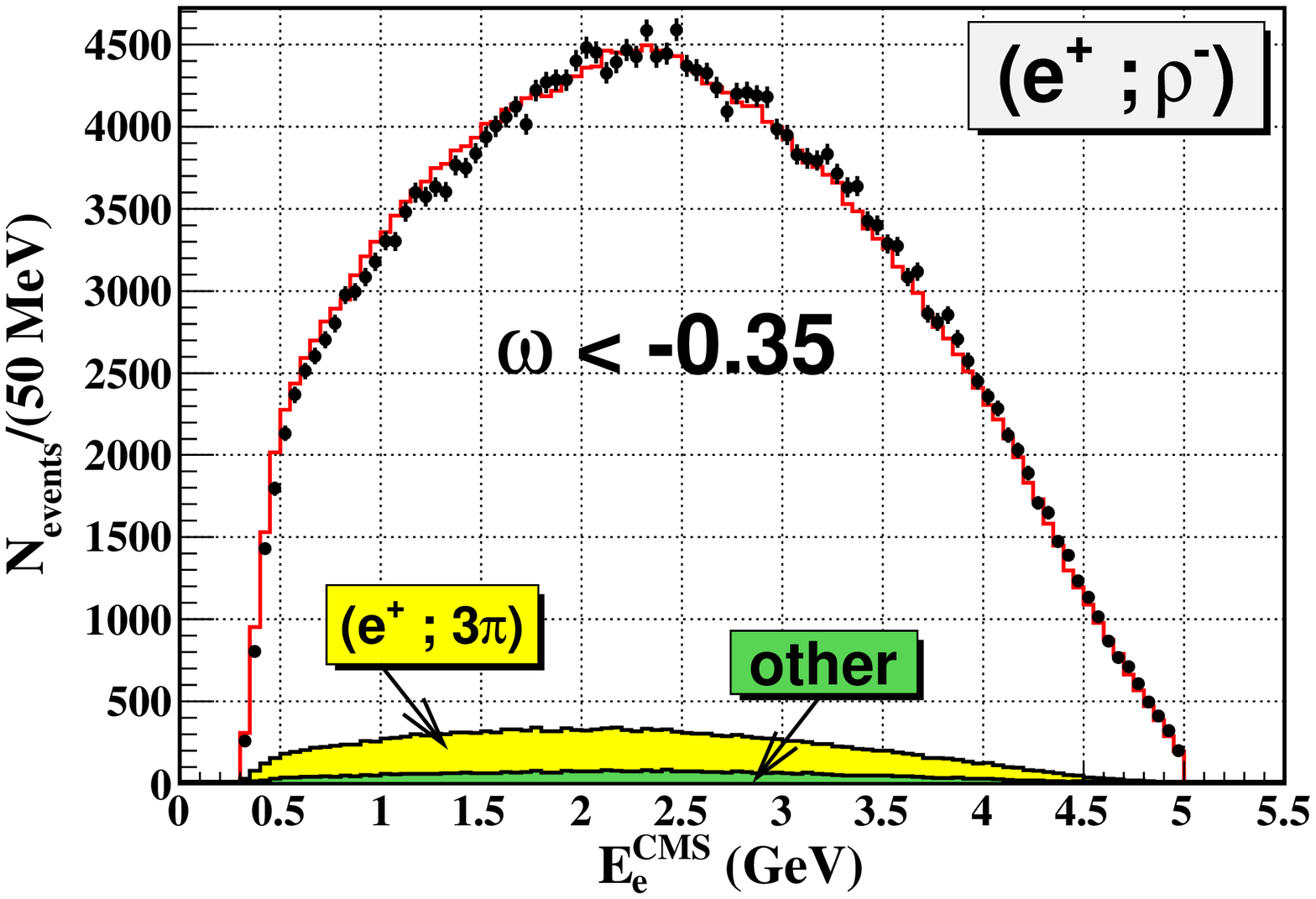} 
\includegraphics[width=0.33\textwidth]{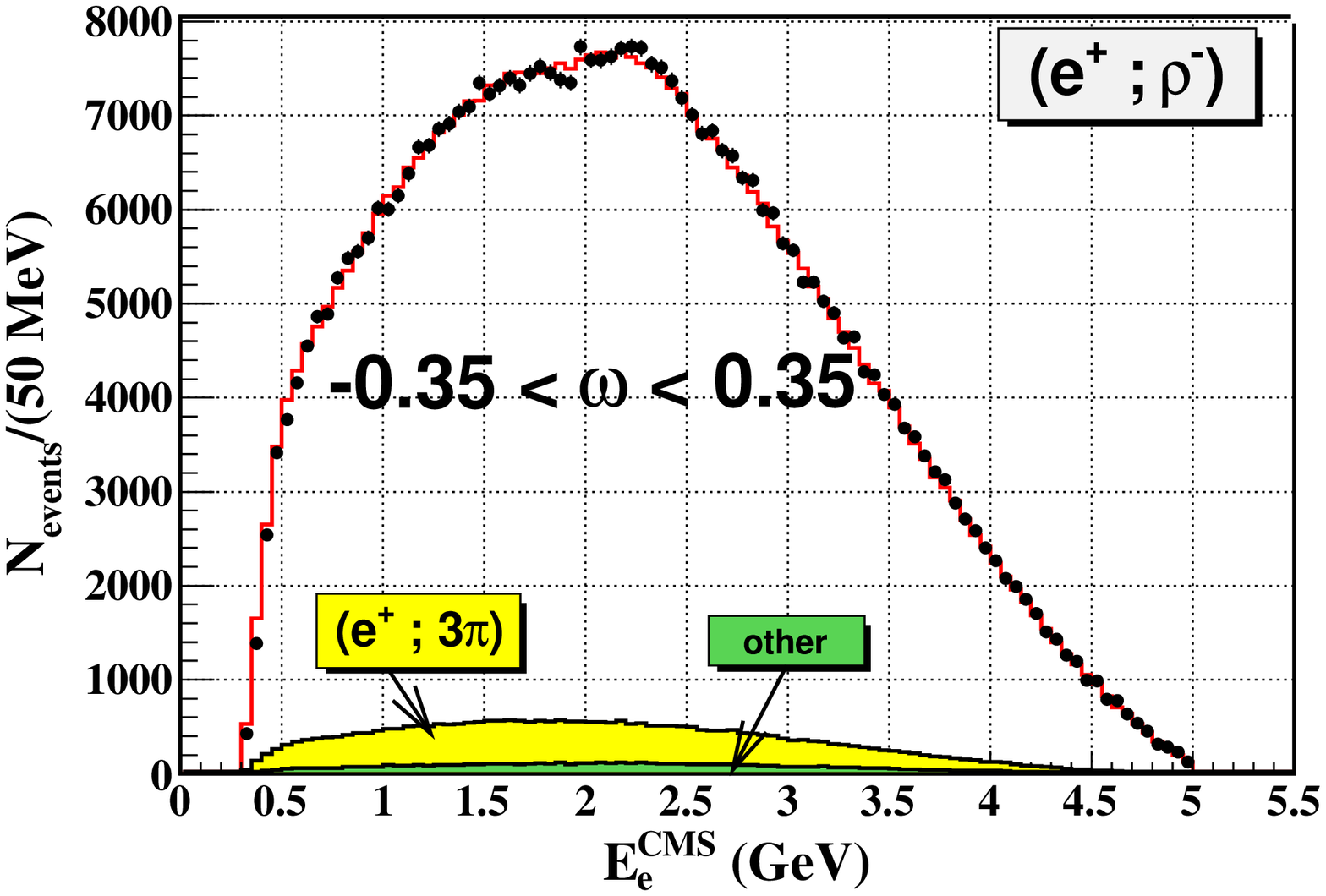} 
\includegraphics[width=0.33\textwidth]{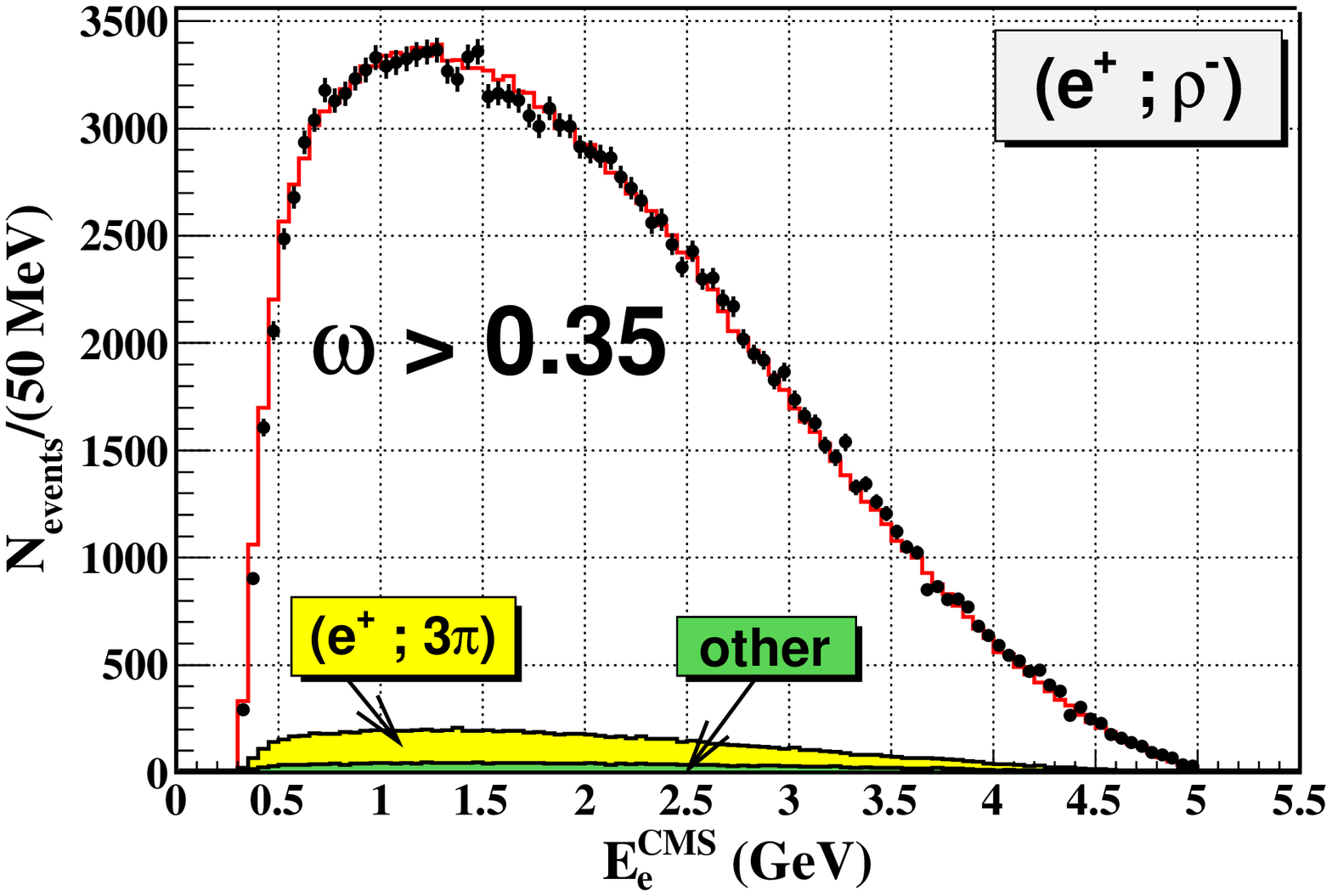} \\ 
\end{tabular}
\caption{Result of the fit of $(e^+,~\rho^-)$ experimental events. 
Three $e^+$ energy spectra are shown for different ranges in $\omega$: 
$\omega < -0.35$ (left), $-0.35 < \omega < 0.35$ (middle), 
$\omega > 0.35$ (right). Points with errors show experimental data, 
histogram - result of the fit. Open histograms show signal events, 
shaded histograms - background contributions.}\label{spincor} 
\end{figure}

The result of the fit of the $(e^+,~\rho^-)$ experimental data is 
illustrated in Fig.~\ref{fitres}. 
Reasonable agreement can be observed for the whole energy range, 
although the relative difference between these spectra indicates a 
remaining systematic effect of about a few percent. The distribution of the 
likelihood per event demonstrates the acceptable quality of the fit. The distribution 
of the $\tau$ helicity sensitive variable $\omega$ \cite{Davier:1} is also 
shown in Fig.~\ref{fitres}.  A spin-spin correlation of tau leptons is clearly 
demonstrated in Fig.~\ref{spincor} for $(e^+,~\rho^-)$ events; the 
$e^+$ energy spectrum shape changes notably as $\omega$ varies 
from $-1$ to $+1$. 

It is confirmed that the uncertainties arising from the physical 
and apparatus corrections to the p.d.f. are well below 1\%; see
Table~\ref{systab}. The statistical uncertainties of the normalisation 
coefficients are kept as small as possible. The contribution to the 
systematic uncertainties of the Michel parameters due to the finite accuracy of the normalisation 
coefficients shown in Table~\ref{systab} are evaluated with the 
entire available generic $\tau^+\tau^-$ MC sample; they 
provide the dominant contributions. We observe a correlation of about
$92$\% between the $\rho$ and $\eta$ parameters. The slope of the 
corresponding error ellipse exhibits an approximate dependence of 
$\Delta\eta\approx 4\Delta\rho$, which is incorporated as an inflated 
uncertainty of the $\eta$ parameter in Table~\ref{systab}. 
\begin{table}[htbp] 
\centering
\caption{Systematic uncertainties of Michel parameters related 
to physical and apparatus corrections, and accuracy of the normalisation  
coefficients ${\cal N}_i$. Values are shown in units of percent 
(i.e. absolute deviation of the Michel parameter is multiplied by 100\%).}\label{systab} 
\begin{tabular}{cc@{~~~}c@{~~~}c@{~~~}c}
\hline
\hline 
Source & $\sigma(\rho)$,\,\% & $\sigma(\eta)$,\,\% & $\sigma(\xi_\rho\xi)$,\,\% & $\sigma(\xi_\rho\xi\delta)$,\,\% \\ 
\hline 
\multicolumn{5}{c}{Physical corrections} \\
\hline
ISR+${\cal O}(\alpha^3)$  & $0.10$  & $0.30$ & $0.20$  & $0.15$ \\ 

$\tau\to\ell\nu\nu\gamma$ & $0.03$  & $0.10$ & $0.09$  & $0.08$ \\ 

$\tau\to\rho\nu\gamma$    & $0.06$  & $0.16$ & $0.11$  & $0.02$ \\ 
\hline 
\multicolumn{5}{c}{Apparatus corrections} \\ 
\hline 
Resolution\,$\oplus$\,brems. & $0.10$ & $0.33$ & $0.11$ & $0.19$ \\

$\sigma(E_{\rm beam})$ & $0.07$ & $0.25$ & $0.03$ & $0.15$ \\ 
\hline 
\multicolumn{5}{c}{Normalisation} \\ 
\hline
$\Delta{\cal N}_1$ &  $0.21$ &  $0.60$ & $0.14$ & $0.12$ \\  
$\Delta{\cal N}_3$ & $<0.01$ & $<0.01$ & $0.35$ & $0.03$ \\ 
$\Delta{\cal N}_4$ &  $0.02$ &  $0.03$ & $0.05$ & $0.23$ \\ 
\hline 
\hline 
Total  & $0.27$ & $0.81$ & $0.47$ & $0.40$ \\ 
\hline
\hline
\end{tabular}
\end{table}

However, we still observe a systematic bias of the order of a 
few percent, especially in the $\xi_\rho\xi$ and $\xi_\rho\xi\delta$ 
Michel parameters. This bias originates from the remaining
inaccuracies in the description of the $\ell-3\pi$ background. 

\section{Summary}

We present a study of Michel parameters in leptonic $\tau$ decays 
using a $485~{\rm fb}^{-1}$ data sample collected at Belle. 
Michel parameters are extracted in the unbinned maximum likelihood 
fit of the $\ell-\rho$ events in the full nine-dimensional phase 
space. We exploit the spin-spin 
correlation of tau leptons to extract $\xi_{\rho}\xi$ 
and $\xi_{\rho}\xi\delta$ in addition to the $\rho$ and
$\eta$ Michel parameters. Although systematic uncertainties coming 
from the physical and apparatus corrections as well as from the 
normalisation are below 1\%, currently we still have a relatively 
large systematic bias in the $\xi_\rho\xi$ and $\xi_\rho\xi\delta$ 
parameters, which originates from the inaccurate description of the 
dominant $\ell-3\pi$ background. 

\section{Acknowledgments}

We thank the KEKB group for the excellent operation of the
accelerator; the KEK cryogenics group for the efficient
operation of the solenoid; and the KEK computer group,
the National Institute of Informatics, and the 
PNNL/EMSL computing group for valuable computing
and SINET4 network support.  We acknowledge support from
the Ministry of Education, Culture, Sports, Science, and
Technology (MEXT) of Japan, the Japan Society for the 
Promotion of Science (JSPS), and the Tau-Lepton Physics 
Research Center of Nagoya University; 
the Australian Research Council and the Australian 
Department of Industry, Innovation, Science and Research;
Austrian Science Fund under Grant No.~P 22742-N16 and P 26794-N20;
the National Natural Science Foundation of China under Contracts 
No.~10575109, No.~10775142, No.~10825524, No.~10875115, No.~10935008 
and No.~11175187; 
the Ministry of Education, Youth and Sports of the Czech
Republic under Contract No.~LG14034;
the Carl Zeiss Foundation, the Deutsche Forschungsgemeinschaft
and the VolkswagenStiftung;
the Department of Science and Technology of India; 
the Istituto Nazionale di Fisica Nucleare of Italy; 
National Research Foundation of Korea Grants
No.~2011-0029457, No.~2012-0008143, No.~2012R1A1A2008330, 
No.~2013R1A1A3007772, No.~2014R1A2A2A01005286, No.~2014R1A2A2A01002734, 
No.~2014R1A1A2006456;
the BRL program under NRF Grant No.~KRF-2011-0020333, No.~KRF-2011-0021196,
Center for Korean J-PARC Users, No.~NRF-2013K1A3A7A06056592; the BK21
Plus program and the GSDC of the Korea Institute of Science and
Technology Information;
the Polish Ministry of Science and Higher Education and 
the National Science Center;
the Ministry of Education and Science of the Russian
Federation and the Russian Federal Agency for Atomic Energy;
the Slovenian Research Agency;
the Basque Foundation for Science (IKERBASQUE) and the UPV/EHU under 
program UFI 11/55;
the Swiss National Science Foundation; the National Science Council
and the Ministry of Education of Taiwan; and the U.S.\
Department of Energy and the National Science Foundation.
This work is supported by a Grant-in-Aid from MEXT for 
Science Research in a Priority Area (``New Development of 
Flavor Physics'') and from JSPS for Creative Scientific 
Research (``Evolution of Tau-lepton Physics'').

\end{document}